\documentclass[floats,prb,aps]{revtex4}
\usepackage{epsfig}

\begin{document}
\title{\bf $ $ \\ About the screening of the charge of a proton migrating in a metal}

\author{A. Lodder}

\affiliation{Faculty of Sciences/Natuurkunde en Sterrenkunde, Vrije Universiteit De
Boelelaan 1081, \\ 1081 HV Amsterdam, The Netherlands}

\date{\today}

\begin{abstract}
\normalsize{
The amount of screening of a  proton in a metal, migrating under
the influence of an applied electric field, is calculated
using different theoretical formulations. First the lowest order screening
expression derived by Sham\cite{sham} is evaluated. In addition 'exact' expressions
are evaluated which were derived according to different approaches.
For a proton in a metal modeled as a jellium the screening
appears to be $15\pm 10\%$, which is neither negligible not
reconcilable with the controversial full-screening point of view
of Bosvieux and Friedel\cite{friedel}. In reconsidering the theory of
electromigration, a new simplified linear-response
expression for the driving force is shown to lead to essentially the
same result as found by Sorbello\cite{sor85}, who has used a rather complicated technique.
The expressions allow
for a reduction such that only the scattering phase shifts of the
migrating impurity are required. 

Finally it is shown that the starting formula
for the driving force of Bosvieux and Friedel leads
exactly to the zero-temperature limit of well-established linear
response descriptions, by which the sting of the controversy
has been removed.
}
\end{abstract}


\maketitle

\section{Introduction}

The amount of screening of a hydrogen atom in a current carrying metal
has been the subject of a long-standing controversy. In brief, considering
the driving force ${\bf F}$ on such an atom as being
composed of two contributions, a direct force
${\bf F}_{\rm d}$ due to the charge of the proton and a wind force
${\bf F}_{\rm w}$ due to transfer of momentum of the current carrying
electrons to the proton, so
\begin{equation}
\label{eq:Feff}
{\bf F}={\bf F}_{\rm d}+{\bf F}_{\rm w}=
(Z_{\rm d}+Z_{\rm w})~e{\bf E}=Z^{*}e{\bf E},
\end{equation}
Bosvieux and Friedel\cite{friedel} found a complete cancellation of ${\bf F}_{\rm d}$,
implying full sceening of the proton charge and only a wind force being operative,
while most other researchers in the field were in favor of at most a very
limited screening.\cite{EL94} According to the convention
in electromigration theory, the forces in Eq. (\ref{eq:Feff})
are written as being proportional 
to corresponding valences and the applied electric field {\bf E}.
The effective valence $Z^{*}$ is the measurable quantity. The wind force
has been calculated
reliably for many systems with {\it ab initio} methods
for the electronic structure. This has been done not only for
migration of interstitials such as hydrogen, but also for substitutional impurities,
including self-electromigration.\cite{EL94,dekkerFCC,dekkerBCC,dekkerJAP,sor97}

In view of the long history of the controversy regarding the
direct force we mention just a few key papers. The linear response
expression for the driving force derived by Kumar and Sorbello\cite{kumar}
was considered as a sound starting point for the resolution of the
controversy. From an evaluation to lowest order in the impurity
potential Sham concluded to a negligible screening.\cite{sham} Using an evaluation
up to all orders in the potential, in 1985 Sorbello found a screening of
at most 25\%.\cite{sor85} This result, combined with a measurement of negligible
screening in V(H) and Ta(H) and a screening of the order of 50\% in Nb(H),
led to a consensus.\cite{AdVerGr}
Nevertheless, Friedel
kept defending that only a wind force was operative, the more so as
Turban et al. had given another support for that point of view.\cite{nozieres}
The confusing feature of the latter work is that their starting formula
is a well-established form of a linear response expression. However,
these authors do not evaluate that expression, but they use a
proportionality argument regarding the expression for another physical quantity.
Another support for the full-screening point of view was given by the present
author \cite{lodder1}. However, that result was considered to be valid in the
low temperature limit only \cite{lodder3,clarification}, and this limit is
a rather academic one in view of the relatively high temperatures
at which electromigration experiments are carried out. More recently Ishida
predicted a screening ranging from 0 to 100\%, but his results were
depending sensitively on the electron density of the host metal.\cite{ishida}

We will present a thorough study of the amount of screening. After summarizing
the main ingredients of the linear response description of the driving force
in \S\ref{theorysumm}, this is
done by first evaluating Sham's screening expression numerically
for a number of model potentials representing the impurity. The results,
given in \S\ref{shamnum},
do not support Sham's conclusion of a negligible screening, but they are
in line with Sorbello's result of a screening of 10-30\%.

Secondly, in \S\ref{theorynew} we will present a very simple evaluation of the linear
response description. This evaluation is supplementary to the
evaluation given by Rimbey and Sorbello\cite{rimby,sor85}
and furthermore much more straightforward. The two descriptions are
compared in \S\ref{comparison}. In \S\ref{Zcorrphase} it
appears to be possible to reduce the final expression for the direct force
valency $Z_{\rm d}$ to
a form containing just the scattering phase shifts of the migrating
impurity potential. Numerical results will be presented in \S\ref{numresults},
and compared with Sorbello's results.\cite{sor85}

Finally, in \S\ref{friedelfout} we will show that the starting expression of Bosvieux
and Friedel for the driving force is precisely the zero temperature limit of
well-established linear response expressions. This is found by describing
the switch-on of the electric field properly and by giving credit to the
hermitian property of the Hamiltonian of the unperturbed system.

\section{Linear-response description} 
\label{theorysumm}
 
The linear response expression for the force on an impurity
with chemical valency $Z_i$
at a position ${\bf R}_1$ due to an applied electric field is given by
\begin{equation}
\label{Windplus}
{\bf F}=
Z_i e {\bf E}-ie E_\nu\int_{0}^{\infty}{\rm d}t e^{- at}
Tr\Big\{ \rho (H)\bigg[{\bf F}_{\rm op} (t) , \sum_j r_j^\nu \bigg] \Big\}\equiv
{\bf F}_{\rm d}^{\rm 'bare'}+{\bf F}_{\rm w}^{\rm total}.
\end{equation}
The first term clearly is the direct force on the bare ion.
The cartesian label $\nu$ runs from 1 to 3, the infinitisimally
positive number $a$ represents the adiabatical switch-on of the
electric field represented by the potential
\begin{equation}
\label{deltaVt}
\delta V(t)=
e {\bf E} e^{at}\cdot \Big(\sum_{j} {\bf r}_{j}
-Z_i\sum_{\alpha}{\bf R}_{\alpha}\Big)\equiv
\delta Ve^{at},
\end{equation}
with $j$ running over the electrons and $\alpha$ over the ions, and the operator
$\rho (H)$ is the grandcanonical density depending on the system
Hamiltonian $H$.
The force operator contains the electron-impurity potential
\begin{equation}
\label{Veiexpliciet}
V_{ei}=\sum_{j,\alpha}v({\bf r}_j-{\bf R}_{\alpha})
\equiv\sum_{j,\alpha}v_j^\alpha,
\end{equation}
which is part of the system Hamiltonian, and is given by
\begin{equation}
\label{FopdVdR}
{\bf F}_{\rm op}\equiv -\nabla_{{\bf R}_1} V_{ei}=-\sum_j
\nabla_{{\bf R}_1}v({\bf r}_j-{\bf R}_1)\equiv\sum_j{\bf f}_j^1.
\end{equation}
Its time dependence refers to the Heisenberg representation
\begin{equation}
\label{FHeisen}
{\bf F}_{\rm op} (t)\equiv e^{iHt}{\bf F}_{\rm op}e^{-iHt}.
\end{equation}
It appears that the second term in Eq. (\ref{Windplus}), which is of course supposed
to lead to the wind force, also contains some screening contribution
to the direct force. The controversy has not to do with the fact
that there is a screening contribution in ${\bf F}_{\rm w}^{\rm total}$, but
it is as to the magnitude of that screening contribution that people don't
agree.
The expression published by Kumar and Sorbello \cite{kumar},
\begin{equation}
\label{WindplusKS}
{\bf F}_{\rm w}^{\rm total}=-\frac{i}{a}E_\nu 
\int_{0}^{\infty}{\rm d}t e^{- at}Tr\Big\{ \rho (H)
\bigg[{\bf F}_{\rm op} (t) , J^\nu\bigg] \Big\},
\end{equation}
follows simply and straightforwardly from a partial integration
of Eq. (\ref{Windplus}) with respect to the time. The current vector
is defined as
\begin{equation}
\label{currentJ}
{\bf J}=ie[\sum_j{\bf r}_j,H]= -e\sum_j\frac{{\bf p}_j}{m}=
\sum_j{\bf j}_j.
\end{equation}
 
The driving force (\ref{Windplus}) can be decomposed as follows:
\begin{equation}
\label{WindPlusStat}
{\bf F}= Z_i e {\bf E}+{\bf F}_{\rm w}^{\rm total}=
 Z_i e {\bf E}+{\bf F}_{\rm w}^{\rm scr}+{\bf F}_{\rm w}^{\rm BF}=
(Z_i+Z^{\rm scr}+Z_{\rm w})e {\bf E}=(Z_{\rm d}+Z_{\rm w})e {\bf E},
\end{equation}
containing the result of Bosvieux and Friedel for the wind force
${\bf F}_{\rm w}^{\rm BF}$ and a screening contribution \cite{friedel}. In all
treatments available ${\bf F}_{\rm w}^{\rm BF}$ can be written in its general form
\begin{equation}
\label{eq:Fdens}
{\bf F}_{\rm w}^{\rm BF}=
-\int \delta n({\bf r}) \nabla_{{\bf R}_{1}}v^{1}d^{3}r,
~~~~~~{\rm with}~~~~~v^{1}=v({\bf r}-{\bf R}_1).
\end{equation}
The precise explicit form depends on the level of approximation
used to represent $\delta n({\bf r})$, which is the local deviation
of the electron density from its unperturbed host
value due to the applied field and the presence
of the impurity. From now on we will concentrate on $Z_{\rm d}$.

All previous relevant descriptions have been given for the
electron-impurity system, for which the Hamiltonian $H$ can be
written as a sum of single particle Hamiltonians $h$, so
\begin{equation}
\label{Hh}
H = \sum_j h^{j}~~~~~~~{\rm with}~~~~~~~h=h_0+v=h_0+\sum_\alpha v^\alpha.
\end{equation}
This allows for a reduction of the many body expression in Eq. (\ref{Windplus})
to the following single particle expression,
\begin{equation}
\label{Windred}
{\bf F}_{\rm w}^{\rm total}=-ie E_\nu \int_{0}^{\infty}
{\rm d}t~e^{- at}~tr\Big\{[r^\nu,n(h)]{\bf f}^1 (t)\Big\},
\end{equation}
where $n(h)$ is the Fermi-Dirac distribution function in operator form
\begin{equation}
\label{nhop}
n(h) = \frac{1}{e^{\beta(h - \epsilon_{\rm F})} + 1}.
\end{equation}
It has been shown explicitly that if in the right hand side
of Eq. (\ref{Windred}) the statistical operator is replaced by
this operator for the free particle system, so $n(h)\rightarrow
n(h_0)$, the Bosvieux-Friedel wind force expression ${\bf F}_{\rm w}^{\rm BF}$
arises \cite{lodder2}. That means that the screening part is
given by
\begin{equation}
\label{Fscr}
{\bf F}_{\rm w}^{\rm scr} =-ie E_\nu \int_{0}^{\infty}
{\rm d}t~e^{- at}~tr\Big\{[r^\nu,n(h)-n(h_0)]{\bf f}^1 (t)\Big\}=
Z^{\rm scr}e{\bf E}.
\end{equation}
The screening valency $Z^{\rm scr}$ is defined as
\begin{equation}
\label{Zscr}
Z^{\rm scr}=-\frac{i}{3}\int_{0}^{\infty}
{\rm d}t~e^{- at}~tr\Big\{[{\bf r},n(h)-n(h_0)]\cdot{\bf f}^1 (t)\Big\},
\end{equation}
in which the factor of $\frac{1}{3}$ comes from the fact that all three terms
in the inner product of the vectors ${\bf r}$ and ${\bf f}^1$
contribute equally.
In all further evaluations the metallic host is modeled by a jellium, which is
the only model used so far in the literature for the study of the direct force
problem. This means that the electrons are perturbed by
the random distribution of impurities only.
Following Sham\cite{sham} we now first consider the result
to lowest (second) order in the impurity potential $v$.

\section{Evaluation of Sham's expression}
\label{shamnum}
 
The evaluation of Eq. (\ref{Zscr}) to lowest order in $v$
requires the expansion of the statistical operator $n(h)-n(h_0)$ in $v$,
\begin{equation}
\label{nhmnh0}
n(h)=n(h_0)-n(h)\int_0^{\beta}ds~e^{sh}~ve^{-sh_0}(1-n(h_0)),
\end{equation}
while in the time dependence of ${\bf f}^1$ one can replace $h$ by
$h_0$. One obtains
\begin{equation}
\label{ZscrSham}
Z^{scr}=-\frac{4}{3m}\sum_{kk'}(k^2-{\bf k}\cdot{\bf k}')\frac{|v_{kk'}|^2}
{(\epsilon_{k}-\epsilon_{k'})^2+a^2}\left(\frac{\partial n_k}{\partial \epsilon_k}
-\frac{n_{k}-n_{k'}}{\epsilon_{k}-\epsilon_{k'}}\right),
\end{equation}
where ${\bf k}$ is a free electron wave vector and $\epsilon_k$ is
the corresponding energy. The matrix element $<{\bf k}|[r^\nu,n(h)-n(h_0)]
|{\bf k}'>$ is most easily evaluated if one realizes, that it is
equal to $i(\frac{\partial}{\partial k_\nu}+\frac{\partial}{\partial k_\nu'})
<{\bf k}|n(h)-n(h_0)|{\bf k}'>$.
Following Sham and Sorbello\cite{sor85} the
potential $v$ refers to the migrating impurity only. Sham stored part of
the presence of the impurities through the relacement $a\rightarrow\tau^{-1}$,
$\tau$ being the transport relaxation time due to the impurities,
which can be justified by an average over the distribution of the
impurities in the time dependence of the force operator.
Both Sham and Sorbello were able to make their complete derivations
after taking the $T\rightarrow 0$ limit only. It has been shown that
Eq. (\ref{ZscrSham}) reduces to Sham's expression after taking that limit.
\cite{lodphysica}

A numerical evaluation of $Z^{scr}$ becomes possible if one employs
the spherical wave expansion for a plane wave, converts the summations over
the wave vectors to integrals and carries out the angular integrals
over the directions of the wave vectors. After using the relation between
$k^2$ and the energy $\epsilon_k$ one ends up at
\begin{equation}
\label{ZscrSham3}
Z^{scr}=-\frac{4}{3\pi^2m}\int_0^\infty d\epsilon_{k}\int_0^\infty
d\epsilon_{k'}\frac{\frac{\partial n_k}{\partial \epsilon_k}
-\frac{n_{k}-n_{k'}}{\epsilon_{k}-\epsilon_{k'}}}
{(\epsilon_{k}-\epsilon_{k'})^2+a^2}\sum_{\ell}f_\ell(k,k'),
\end{equation}
in which the function $f_\ell(k,k')$ is defined as
\begin{equation}
\label{fellkkp}
f_\ell(k,k')=\epsilon_k\sqrt{\epsilon_{k'}}v_\ell(k',k)
\bigg[(2\ell+1)kv_\ell(k',k)-2(\ell+1)k'v_{\ell+1}(k',k)\bigg],
\end{equation}
containing the information about the ion potential through
\begin{equation}
\label{vellkpk}
v_\ell(k',k)=\int_0^\infty r^2dr~j_\ell(k'r)v(r)j_\ell(kr).
\end{equation}
The integrand has to be treated with care when $\epsilon_{k'}=
\epsilon_k$, because then the denominator attains the
value $a^2$ which would imply 'singular' behaviour. However,
precisely then the numerator becomes zero,
because $\lim_{\epsilon_{k'}\rightarrow\epsilon_k}
(n_{k}-n_{k'})/(\epsilon_{k}-\epsilon_{k'})\rightarrow
\frac{\partial n_k}{\partial \epsilon_k}$. The crucial part
of the integrand lies in the square around the point
$(\epsilon_{k},\epsilon_{k'})=(\epsilon_{\rm F},\epsilon_{\rm F})$.
In studying the $Z^{scr}$ integral it appears that in that square one has to
keep the Fermi-Dirac distribution function in its finite temperature form.
We could not obtain a reliable stable numerical result by using
Sham's $T\rightarrow 0$ expression. 
\vspace{8mm} 
\begin{figure}[htb]
\epsfig{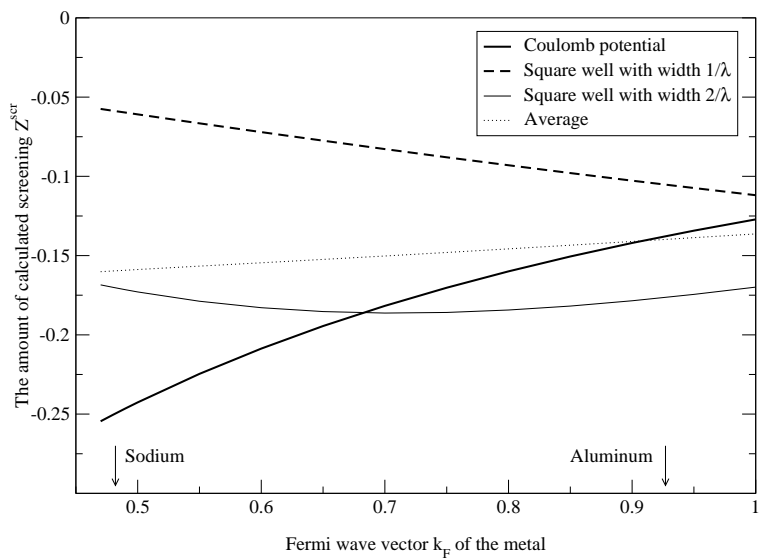}
\caption[]{The amount of screening represented
by $Z^{scr}$ according to Eq. (\ref{ZscrSham3}), for the screened Coulomb
potential and for two square well potentials.}
\label{Zscrfig}
\end{figure}
The result of a numerical evaluation for different ion potentials
is shown in FIG. \ref{Zscrfig}. We used a screened Coulomb potential
\begin{equation}
\label{vScreenedC}
v(r)=-\frac{Z_ie^2e^{-\lambda r}}{r}~~~~{\rm while}~~~~v_{k'k}\equiv\frac{1}{8\pi^3}
\int d^3r~e^{i({\bf k}-{\bf k}')\cdot{\bf r}}v(r),
\end{equation}
with $Z_i=1$ representing a proton in a jellium and the inverse screening
length $\lambda$ given by the Thomas-Fermi expression\cite{ziman}
\begin{equation}
\label{lambda}
\lambda^2=4\pi e^2{\cal{N}}(\epsilon_{\rm F})~~~{\rm with}~~~{\cal N}(\epsilon)=
\frac{m\sqrt{2m\epsilon}}{\pi^2}~\rightarrow~ v_{kk}=
-\frac{Z_i}{8\pi^3{\cal N}(\epsilon_{\rm F})}.
\end{equation}
In addition square well
potentials were employed in the same spirit as Sorbello did.\cite{sor85}
The width $r_0$ of the square well potential was chosen to be equal
to the screening length $1/\lambda$ and twice as large.
The corresponding well depth $v_0$ was determined by the condition
$v_0/v_c=0.999$, where $v_c=\pi^2/8mr_0^2$ is the critical value of the well depth
for which a bound state forms. For further details, see Ref. \cite{sor85}.
The value of $\lambda$ is determined by the Fermi energy. While
Sorbello chose five values for the Fermi energy, typical for metals
ranging from sodium to aluminum, we have done the calculation for
a whole range of Fermi energies. The results are plotted as a function
of the Fermi wave number $k_{\rm F}$. The $k_{\rm F}$ values of
sodium and aluminum are indicated.
 
Because $\lambda$ increases monotonically with the Fermi energy, the
range of the corresponding screened Coulomb potential decreases
with increasing $k_{\rm F}$, which reduction in strength is seen
clearly in the solid curve. 
\begin{figure}[htb]
\epsfig{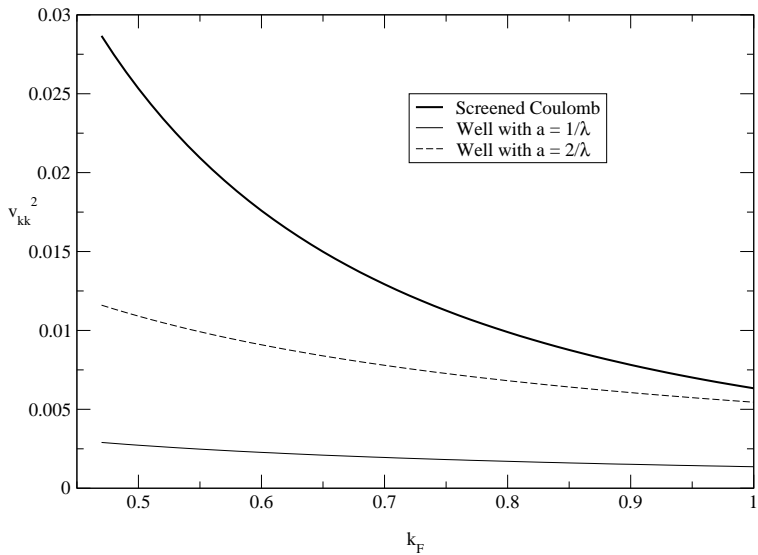}
\caption[]{$v_{kk}^2$ plotted as a function
of $k_{\rm F}$, for the two square well potentials and the
screened Coulomb potential.}
\label{vkk2}
\end{figure}
In FIG. \ref{vkk2} the $v_{kk}^2$ curves are plotted
for the three potentials. A clear decrease is seen for the Coulomb potential,
and a rather flat
behaviour for the square well potentials, while the one with
$2/\lambda$ is markedly stronger than the one with the smaller width.
Apparently, the screening to second order in the impurity potential
is not negligible at all, but on the average as large as
$15\pm10\%$. As a guide for the eye we gave the average
of $Z^{scr}$ for the three potentials as a dotted line.

For security we evaluated an alternative expression for $Z^{scr}$,
given by
\begin{equation}
\label{ZscrShamSF2}
Z^{scr}=-\frac{4}{3m}\sum_{kk'}(k^2-{\bf k}\cdot{\bf k}')|v_{kk'}|^2
\frac{n_{k}-n_{k'}}{\epsilon_{k}-\epsilon_{k'}}
\frac{(\epsilon_{k}-\epsilon_{k'})^2-a^2}
{[(\epsilon_{k}-\epsilon_{k'})^2+a^2]^2}.
\end{equation}
This expression follows if Eq. (\ref{Zscr}) is modified such, that
the dipole operator ${\bf r}$ commutes with the force operator ${\bf f}^1 (t)$
instead of with the statistical operator. While $Z^{scr}$ given by
Eq. (\ref{ZscrSham}) reduces to Sham's expression after taking the
$T\rightarrow 0$ limit, such a proof is not available for $Z^{scr}$ given by
Eq. (\ref{ZscrShamSF2}). On the other hand, the two expressions
(\ref{ZscrSham}) and (\ref{ZscrShamSF2}) are equivalent, being related
to each other through a partial integration for the derivatives
with respect to the ${\bf k}$ and ${\bf k}'$ vectors.
The alternative for Eq. (\ref{ZscrSham3}) becomes
\begin{equation}
\label{ZscrSham3SF2}
Z^{scr}=-\frac{4}{3\pi^2m}\int_0^\infty d\epsilon_{k}\int_0^\infty
d\epsilon_{k'}\frac{n_{k}-n_{k'}}{\epsilon_{k}-\epsilon_{k'}}
\frac{(\epsilon_{k}-\epsilon_{k'})^2-a^2}
{[(\epsilon_{k}-\epsilon_{k'})^2+a^2]^2}\sum_{\ell}f_\ell(k,k').
\end{equation}
If one takes proper care of the higher sensitivity of
the expression (\ref{ZscrSham3SF2}) to the choice of the infinitisimal
parameter $a$ the results turn out to be the same. It can be taken
relatively small, much smaller than a typical value of 0.01 for
the inverse transport relaxation time. In fact, it is the mesh of the integration
that determines the lower limit of $a$. For the expressions (\ref{ZscrSham3})
and (\ref{ZscrSham3SF2}) it was never larger than 0.00015 and 0.005
respectively. On the other hand, for $a=0.01$ the screening represented
by the curves in FIG. \ref{Zscrfig} reduces by at most 2\%.
 
\section{Reduction of linear-response formula for the screening}
\label{theorynew}
 
The evaluation of the linear-response formula (\ref{Zscr}) to all orders
in the impurity potential
can be achieved by restricting the evaluation
to a system with one impurity in a jellium, which is in accordance
with earlier work by others.\cite{sor85}
In that case ${\bf f}^1=i[{\bf p},h]=i[{\bf p},h_0+v^1]$. By
writing Eq. (\ref{Zscr}) in terms of eigenstates
of $h\rightarrow h_0+v^1$, labeled by $q$ and $q'$, one 
can carry out the time integral and finds
\begin{eqnarray}
\label{WindStatq}
&&Z^{scr}
=\frac{i}{3}\sum_{qq'}<q|[{\bf r},n(h)-n(h_0)]|q'>\cdot
\frac{\epsilon_q-\epsilon_{q'}}{\epsilon_{q'}-\epsilon_q+ia}{\bf p}_{q'q}
\nonumber\\&&~~~~~=\frac{i}{3}\sum_{qq'}<q|[{\bf r},n(h)-n(h_0)]|q'>\cdot
\left(-1+\frac{ia}{\epsilon_{q'}-\epsilon_q+ia}\right)
{\bf p}_{q'q}\nonumber=\\&&\hspace{8mm}
-\frac{i}{3} tr\Big\{[{\bf r},n(h)-n(h_0)]\cdot{\bf p}\Big\}
+\frac{i}{3}\sum_{qq'}^{\epsilon_q=\epsilon_{q'}}
<q|[{\bf r},n(h)-n(h_0)]|q'>\cdot{\bf p}_{q'q}.
\end{eqnarray}
The first term in the last line reduces to $-Z_i$
because $i~tr\{[r^\nu,n(h)-n(h_0)]p^\mu\}=\\ i~
tr\{\left(n(h)-n(h_0)\right)[p^\mu,r^\nu]\}=tr\{n(h)-n(h_0)\}
\delta_{\nu,\mu}=Z_i\delta_{\nu,\mu}$. One arrives at
\begin{equation}
\label{correction}
Z^{scr}=-Z_i+Z_{\rm corr}~~~~~~~{\rm with}~~~~~~~
Z_{\rm corr}\equiv\frac{i}{3}\sum_{qq'}^{\epsilon_q=\epsilon_{q'}}
<q|[{\bf r},n(h)-n(h_0)]|q'>\cdot{\bf p}_{q'q}.
\end{equation}
In view of Eq. (\ref{WindPlusStat}) this would imply a correction term
$Z_{\rm corr}=Z_{\rm d}$ to the cancellation of the bare direct valency $Z_i$.
The step of subtracting and adding an $ia$ term in the
numerator in the second line of Eq. (\ref{WindStatq}) may
look somewhat artificial, and the $ia$ factor
creates the impression to lead to a zero result in the $a\rightarrow 0$
limit. But in following Sorbello it is seen in the third line that for the
$\epsilon_q=\epsilon_{q'}$ terms the $ia$ factor cancels,
and the remaining terms give a finite contribution.\cite{sorbello0}

Sorbello \cite{sor85} starts from a result obtained by Rimbey and Sorbello
\cite{rimby} through an evaluation of Eq. (\ref{WindplusKS})
and finds after some rewritings for $Z_{\rm d}$
\begin{equation}
\label{Sorb7op}
Z_{\rm d}=-\frac{2}{3\pi m}{\rm Im}tr\Big\{p^2\left(G(\epsilon_{\rm F})-
G^0(\epsilon_{\rm F})\right)\Big\}.
\end{equation}
The single particle Green's function $G(\epsilon)$
for one impurity in a jellium, with $h=h_0+v^1$, and the free electron
Green's function  $G^0(\epsilon)$ are given by
\begin{equation}
\label{GG0} 
G(\epsilon)=\frac{1}{\epsilon +ia-h}~~~~{\rm and}~~~~
G^0(\epsilon)=\frac{1}{\epsilon +ia-h_0}~~~~{\rm with}~~~~h_0=\frac{p^2}{2m}.
\end{equation}
This form for $Z_{\rm d}$ is Sorbello's Eq. (12) and it
implies a cancellation of $Z_i$ present in his Eq. (7).
In order to distinguish our result for $Z_{\rm d}$ from
Sorbello's $Z_{\rm d}$ we keep the notation $Z_{\rm corr}$.

\section{Comparison of the two descriptions}
\label{comparison}

While the description by Rimbey and Sorbello is rather involved and
the result (\ref{correction}) is obtained in a few lines, it is worth while
to compare the final expressions. 
We first evaluate $Z_{\rm corr}$ to lowest order in the impurity potential,
for which we take a screened Coulomb potential, Eq. (\ref{vScreenedC}).
By using Eq. (\ref{nhmnh0})
and the equality
\begin{equation}
\label{lambdaform}
\bigg[{\bf r},n(h)\bigg]=-\frac{i}{m}\int_0^\beta ds~
n(h)e^{sh}{\bf p} e^{-sh}\left(1-n(h)\right),
\end{equation}
in which the electron mass $m=\frac{1}{2}$ in atomic units,
one finds straightforwardly for $Z_{\rm corr}^{0}$
\begin{equation}
\label{Zcorr0}
Z_{\rm corr}^{0}=\frac{i}{3}\sum_{k}
<k|[{\bf r},n(h)-n(h_0)]|k>\cdot{\bf k}=-4\pi\sqrt{\epsilon_{\rm F}}v_{kk}=Z_i
\equiv Z_i({\rm pot}),
\end{equation}
in which a quantity $Z_i({\rm pot})=-4\pi\sqrt{\epsilon_{\rm F}}v_{kk}$ is
defined to be used below. For the screened Coulomb potential this
quantity is equal to $Z_i=1$, but this is not the case for other potentials.
This result from an explicit calculation follows also if one
writes the sum over the free space states $|k>$ as a trace and
uses the equality given in the sentence just below Eq. (\ref{WindStatq}).

Similarly one finds for Sorbello's $Z_{\rm d}$ to lowest order
in the impurity potential, writing the trace in Eq. (\ref{Sorb7op})
in terms of free space states labeled by $k$,
\begin{equation}
\label{ZdSor0}
Z_d^{(0)}=-\frac{2}{3\pi m}{\rm Im}\int d^3k~k^2
G^0_k(\epsilon_{\rm F})v_{kk}G^0_k(\epsilon_{\rm F})=
-4\pi\sqrt{\epsilon_{\rm F}}v_{kk}=Z_i.
\end{equation}
This is obtained by using the following two equalities,
\begin{equation} 
\label{dGdEImG}
\frac{\partial}{\partial\epsilon_{k}}G_{k}^0(\epsilon)=\frac{\partial}
{\partial\epsilon_{k}}\frac{1}{\epsilon-\epsilon_{k}+ia}=
G_{k}^0(\epsilon)G_{k}^0(\epsilon)
~~~~~{\rm and}~~~~~\lim_{a\rightarrow 0}{\rm Im}G^0_k(\epsilon_{\rm F})=
-\pi \delta(\epsilon_{\rm F}-\epsilon_{k}).
\end{equation}

Apparently, to lowest order in the impurity potential the two
final expressions $Z_{\rm corr}$ and $Z_{\rm d}$ are equal and they reproduce the
bare valency of the migrating ion. Although the complete expressions
are not equal, an 'almost' equality can be derived.
We rewrite $Z_{\rm corr}$ by applying Eq. (\ref{lambdaform}) both for
$h$ and $h_0$. After inserting a complete set of free electron states
in the $h_0$ term in Eq. (\ref{correction})
and carrying out the integral over $s$, one finds
\begin{eqnarray}
\label{Contact1}
&&Z_{\rm corr}=
\frac{\beta}{3m}\sum_{qq'}^{\epsilon_q=\epsilon_{q'}}\Big(
n_q(1-n_q){\bf p}_{qq'}-\sum_{k''}n_{k''}(1-n_{k''})<q|k''>{\bf k}''<k''|q'>\Big)
\cdot{\bf p}_{q'q}\nonumber\\&&\rightarrow
-\frac{1}{3\pi m}\sum_{qq'}^{\epsilon_q=\epsilon_{q'}}\Big(
{\rm Im}G_q(\epsilon_{\rm F}){\bf p}_{qq'}-\sum_{k''}{\rm Im}G_{k''}^0(\epsilon_{\rm F})
<q|k''>{\bf k}''<k''|q'>\Big)\cdot{\bf p}_{q'q}\nonumber\\&&=
-\frac{1}{3\pi m}{\rm Im}\sum_{qq'}^{\epsilon_q=\epsilon_{q'}}<q|{\bf p}
\left(G(\epsilon_{\rm F})-G^0(\epsilon_{\rm F})\right)|q'>\cdot{\bf p}_{q'q}.
\end{eqnarray}
In the transition from the first to the second line the $T\rightarrow 0$
limit was taken, for which $\beta n_q(1-n_q)=-\frac{\partial}{\partial\epsilon_q}
n_q\rightarrow\delta(\epsilon_q-\epsilon_{\rm F})=-\frac{1}{\pi}
{\rm Im}G_q(\epsilon_{\rm F})$. Both Sham and Sorbello give their
elaborated expressions in this $T\rightarrow 0$ limit. The similarity
of this last line with Sorbello's $Z_{\rm d}$, Eq. (\ref{Sorb7op}),
is striking. The factor of 2 reflects whether the electron spin
degeneracy has been accounted for explicitly or not. In fact, if in the last
line of Eq. (\ref{Contact1}) the states $q$ and $q'$ are replaced by the
unperturbed ones $k$ and $k'$, it reduces to Sorbello's expression.
This implies an intriguing equality indeed, and it shows that the two
descriptions are closely related.
 
\section{The correction term in terms of phase shifts} 
\label{Zcorrphase}
 
For the evaluation of the correction term $Z_{\rm corr}$ as it
is defined in Eq. (\ref{correction}) one needs the states $|q>$.
These states are the eigenstates of a system with one impurity
in free space. It is known that the scattering states $|\psi_k>$
for this system, which have
a one-to-one correspondence to the free space states $|k>$, are exact solutions of
the Schr\"{o}dinger for one impurity in free space as well. It appears that
the evaluation of $Z_{\rm corr}$ becomes relatively simple if one uses the
scattering states instead of the true eigenstates. We return to this point below.

The expansion of the scattering state $<{\bf r}|\psi_k>\equiv\psi_k ({\bf r})$
in terms of spherical harmonics is given by
\begin{equation}
\label{psikrL}
\psi_k ({\bf r})=\frac{4\pi}{\sqrt\Omega}\sum_L
i^\ell Y_L^\ast(\hat{k})R_\ell(r,k)Y_L(\hat{r}).
\end{equation}
The angular momentum label $L$ combines the labels $\ell$
and $m$, so $L\equiv \ell m$, and $R_\ell(r,k)$ is the radial solution
of the Schr\"{o}dinger equation at the energy $\epsilon_k$
for a spherically symmetric potential $v$ centered at
the origin. For $r$ outside the range of the potential $R_\ell(r,k)$
can be written in terms of  the scattering $t$ matrix $t_\ell =-\frac{1}{k}\sin\delta_\ell
\exp(i\delta_\ell)$ as $j_\ell(kr)-ikt_\ell h^+_\ell(kr)$, where
$\delta_\ell$ are the phase shifts.
This means that for a plane wave $R_\ell(r,k)\rightarrow j_\ell(kr)$.
The box normalization
in the system volume $\Omega$ induces a discrete set of $k$ values.
In the properties to be presented below a delta function normalization
will be used, which means that in Eq. (\ref{psikrL}) the system volume $\Omega$
has to be replaced by $8\pi^3$.
Using the expansion Eq. (\ref{nhmnh0}), the equality $<k'|v|\psi_k>=
t_{k'k}$ which holds for scattering states $|q>\rightarrow|\psi_k>$, and
the overlap property for scattering states
\begin{equation}
\label{kacpsik}
<k'|\psi_k>=\frac{\delta(k-k')}{k^2}\sum_L Y_L(\hat{k}')Y_L^\ast(\hat{k})
\bigg(1-ikt_\ell\bigg),
\end{equation}
one finds for $Z_{\rm corr}$
\begin{equation}
\label{ZcorrPhase}
Z_{\rm corr}=-\frac{4}{3\pi m}
\int_0^\infty k^3dk\Big(\frac{\partial}{\partial\epsilon_k}
\delta(\epsilon_k-\epsilon_{\rm F})\Big)F(\epsilon_k)=
\frac{2}{\pi}\left(F(\epsilon_{\rm F})+\frac{2}{3}\epsilon_{\rm F}^
{\frac{3}{2}}\frac{\partial}{\partial\epsilon_{\rm F}}\bar{F}(\epsilon_{\rm F})\right),
\end{equation}
in which the function $F(\epsilon_k)$ is given by
\begin{equation}
\label{FEk}
F(\epsilon_k)=\frac{1}{4}\sum_{\ell}(\ell+1)\bigg(\sin2\delta_{\ell}+
\sin2\delta_{\ell+1}\bigg)
\bigg(\cos^2(\delta_{\ell}-\delta_{\ell+1})+1\bigg),
\end{equation}
and $\bar{F}(\epsilon_k)\equiv\frac{1}{k}F(\epsilon_k)$.
Crucial steps of the derivation of Eq. (\ref{ZcorrPhase}) are given in Appendix
\ref{crucial}. This expression can be evaluated simply, because it is just
a function of the phase shifts of the impurity potential at the Fermi energy.
$Z_{\rm corr}$ as it is given by Eq. (\ref{ZcorrPhase}) has to
be compared with the lowest order expression, obtained by the replacements
$|q>\rightarrow|k>$ and $|q'>\rightarrow|k'>$. For the sake of a proper
comparison this expression has to be evaluated in a similar way, by
the use of scattering states. This way one obtains
\begin{equation}
\label{ZcorrPhase01}
Z_{\rm corr}^{0}=-\frac{4}{3\pi m}\int_0^\infty k^3dk
\Big(\frac{\partial}{\partial\epsilon_k}\delta(\epsilon_k-\epsilon_{\rm F})\Big)
F^{0}(\epsilon_k)=\frac{2}{\pi}\left(F^{0}(\epsilon_{\rm F})+\frac{2}{3}
\epsilon_{\rm F}^{\frac{3}{2}}\frac{\partial}{\partial\epsilon_{\rm F}}
\bar{F}^{0}(\epsilon_{\rm F})\right),
\end{equation}   
in which the function $F^{0}(\epsilon_k)$ is given by 
\begin{equation} 
\label{F0Ek}
F^{0}(\epsilon_k)=\frac{1}{2}\sum_{\ell}(2\ell+1)\sin2\delta_{\ell}
=\sum_{\ell}(2\ell+1)\sin\delta_{\ell}\cos\delta_{\ell},
\end{equation}   
and $\bar{F}^{0}(\epsilon_k)\equiv\frac{1}{k}F^{0}(\epsilon_k)$.
The right-hand sides of Eqs. (\ref{ZcorrPhase01}) and (\ref{Zcorr0})
can have different numerical values, because the two elaborations
are different in character. Comparison of these numerical values
can be considered as a test of the error made in using scattering states instead
of the true eigenstates.
Another test of this error is obtained by evaluating the simplified
lowest order result for $Z_{\rm corr}$ explicitly, to be denoted as
$Z_{\rm corr}^{00}$, using the expansion (\ref{nhmnh0}). One finds
\begin{equation} 
\label{ZcorrPhase02}
Z_{\rm corr}^{00}=tr\{n(h)-n(h_0)\}=\frac{2}{\pi}
\sum_\ell(2\ell+1)\sin\delta_{\ell}\cos\delta_{\ell}=
\frac{2}{\pi}F^{0}(\epsilon_k)\neq
\frac{2}{\pi}\sum_\ell(2\ell+1)\delta_{\ell}=Z_{\rm F},
\end{equation}
in which $Z_{\rm F}$ stands for the Friedel sum.
The inequality in Eq. (\ref{ZcorrPhase02}) must be attributed to
the use of scattering states instead of the true eigenstates.
The difference between these two types of states has been
stressed by Fenton \cite{fenton} and commented on by the
present author \cite{lodderselfen}. A scattering state is prepared
such that it is an unperturbed state at $t=-\infty$, which develops under
the influence of a scattering potential located at some position in the
system. An eigenstate has to be constructed using the boundary
conditions of the system in addition to the properties of
the potential. In terms of scattering theory eigenstates contain
the influence of backscattering by the boundaries in addition to
the information about the scattering by the potential. It is clear
that the third member of Eq. (\ref{ZcorrPhase02}) reduces to
$Z_{\rm F}$ in the small phase-shifts limit
$\delta_{\ell}\rightarrow 0$.

 
Sorbello's equation for $Z_{\rm d}$, Eq. (\ref{Sorb7op}),
can be evaluated through the use of scattering states as well.
We first give the form corresponding to the eigenstates of
$h$, denoted as usual by $|q>$, followed by the result obtained
by using scattering states.
\begin{eqnarray}
\label{SorbScatt}
Z_d&=&-\frac{2}{3\pi m}{\rm Im}
\sum_q\sum_{k'}<q|k'>{k'}^2G^0_{k'}(\epsilon_{\rm F})<k'|v|q>G_q(\epsilon_{\rm F})
\nonumber\\ &\rightarrow&\frac{4}{3\pi}\frac{\partial}{\partial \epsilon_{\rm F}}
\int_0^\infty\epsilon_k^{\frac{3}{2}}
d\epsilon_k\delta(\epsilon_{\rm F} -\epsilon_k)\bar{F}^{0}(\epsilon_k)=
\frac{4}{3\pi}\frac{\partial}{\partial \epsilon_{\rm F}}\epsilon_{\rm F}^{\frac{3}{2}}
\bar{F}^{0}(\epsilon_{\rm F}).
\end{eqnarray}
As above, for scattering states the potential matrix element becomes equal to
the $t$ matrix element $t_{k'k}$. The energies $\epsilon_k$
and $\epsilon_q$ are equal, being connected by a delta function as
shown in Eq. (\ref{kacpsik}).
Interestingly, this rewritten $Z_d$ is
equal to our $Z_{\rm corr}^{0}$ given  by Eq. (\ref{ZcorrPhase01}).
This again shows the close relationship between the results obtained
through the simplified approach presented here and Sorbello's results.

Sorbello calculated $Z_d$ of Eq. (\ref{Sorb7op}) using the square well
potentials described in \S\ref{shamnum}. To that end he derives his rewritten form
Eq. (17), which we reproduce in a slightly different notation as
\begin{eqnarray} 
\label{Sorb7opform1}
Z_d&=&
\frac{4v_0k_{\rm F}}{3\pi}\int_0^{r_0}r^2dr\sum_\ell(2\ell+1)
\left(R_\ell^2(r,k_{\rm F})-j_\ell^2(k_{\rm F}r)\right)
\nonumber\\ &&+\frac{4\epsilon_{\rm F}}{3\pi}
\sum_\ell(2\ell+1)\frac{\partial\delta_\ell}{\partial\epsilon}
\rfloor_{\epsilon_{\rm F}}+\frac{4k_{\rm F}v_0r_0^3}{9\pi}.
\end{eqnarray}
In the derivation has been used, that $\frac{p^2}{2m}=h-v=h-\epsilon_{\rm F}+
\epsilon_{\rm F}-v$, $G-G^0=GvG^0$, $(h-\epsilon_{\rm F})G=-1$,
$-{\rm Im}Tr\{G-G^0\}=\sum_\ell(2\ell+1)\frac{\partial\delta_\ell}
{\partial\epsilon}\rfloor_{\epsilon_{\rm F}}$, ${\rm Im}G^0({\bf r},{\bf r})=
-k_{\rm F}\sum_Lj_L^2({\bf r})$, and ${\rm Im}G({\bf r},{\bf r})=
-k_{\rm F}\sum_LR_L^2({\bf r})$, with $R_L({\bf r})\equiv R_\ell
(r,k_{\rm F})Y_L(\hat{r})$. For $r\leq r_0$ the radial solution
$R_\ell(r,k_{\rm F})\propto j_\ell(k_{v}r)$, with
$k_{v}=\sqrt{k_{\rm F}^2+v_0}$.
For reasons of a proper comparison we evaluated $Z_d$ according
to Eq. (\ref{Sorb7opform1}) up to $\ell =2$, because Sorbello
restricted himself to $\ell_{\rm max}=0$.
\begin{figure}[htb]
\vspace{15mm}
\epsfig{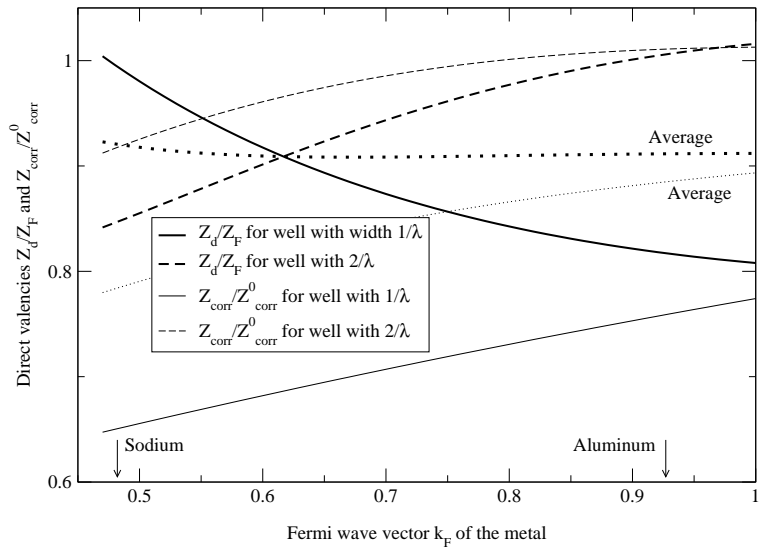}
\caption[]{$Z_d/Z_{\rm F}$ and $Z_{\rm corr}/Z_{\rm corr}^{0}$
plotted as a function
of $k_{\rm F}$, for the two square well potentials.}
\label{PaZdcorr}
\end{figure}

\section{Numerical results}\label{numresults} 
 
The expressions obtained will be evaluated for the same
square well model potentials as were used in \S\ref{shamnum}, in that
employing a slight generalization of the  potentials used by Sorbello \cite{sor85}.
Results for $Z_d/Z_{\rm F}$ and $Z_{\rm corr}/Z_{\rm corr}^{0}$
are shown in FIG. \ref{PaZdcorr}. Because the Friedel sum for the
model square well potentials is rarely equal to unity, the use of ratios
gives the proper measure for the screening, in which we follow
Sorbello.

It is seen that for the stronger potentials, with $r_0=2/\lambda$,
Sorbello's boldly dashed curve lies somewhat lower than the dashed
curve for the present description, while they display almost equal
results for the higher $k_{\rm F}$ values. The curves for the weaker
potentials lie lower than those for $r_0=2/\lambda$. Although they
differ considerably for smaller $k_{\rm F}$ values, the curves approach
each other for higher $k_{\rm F}$. Both results imply a decrease of the
inaccuracy related to the $|q>\rightarrow|\psi_k>$ replacement
in the present description for states with increasing $k$ values.
This is reasonable, because larger $k$ values correspond to smaller
wave lengths, which probe the scattering potential more precisely, while
the boundary effect decreases. As a guide for the eye the average
for the two well widths are drawn as dotted lines. From the present description
one comes to a direct valency of $0.85\pm 0.15$ on the average,
while this is $0.91\pm 0.10$ for Sorbello's description. 
So it appears that the $|q>\rightarrow|\psi_k>$
replacement is not too crude in determining a measure for
the amount of screening. The screening
mentioned by Sorbello is based mainly on the $r_0=1/\lambda$ potentials,
because for these potentials his restriction to $\ell_{\rm max}=0$ is reasonable.
If we correct for the higher $\ell$ values we find 0.82 for
$Z_d/Z_{\rm F}$ in aluminum instead of his 0.75. He used the latter value
in mentioning a screening of 25\%. For the sake of completeness
we remark, that if one would compare the boldly dashed curve with
the values given in Sorbello's Table II one would observe considerable
differences. This is due to the fact that for the $r_0=2/\lambda$
potentials the $\ell=1$ and 2 terms contribute significantly.
Taking everything together the
available models and descriptions end up at a screening
between 5 and 30\%. Comparing with the largely metallic-density
dependent result of Ishida, covering the entire range of no screening
to complete screening, the present result can be considered as rather
conclusive, in that complete screening is excluded.\cite{ishida}
\begin{figure}[htb]
\vspace{15mm} 
\epsfig{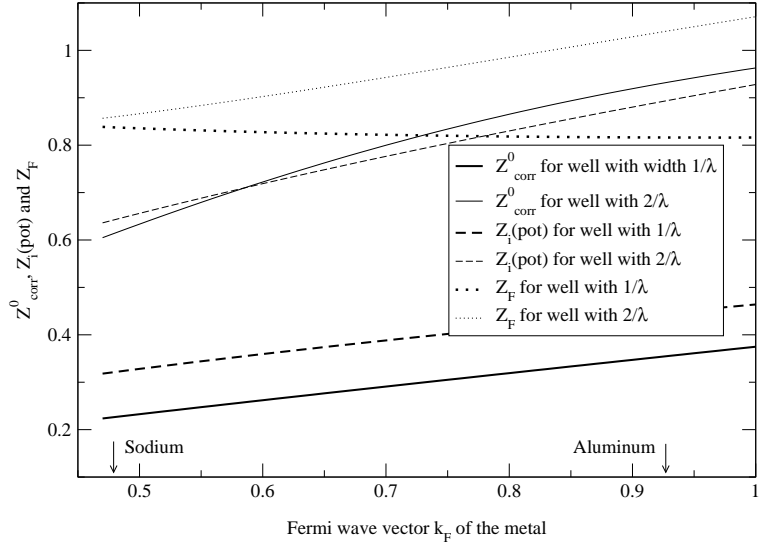}
\caption[]{$Z_{\rm corr}^{0}$, $Z_i(\rm pot)$ and $Z_{\rm F}$ plotted as a function
of $k_{\rm F}$, for the two square well potentials.}
\label{PaZiZ0corr}
\end{figure}

\begin{figure}[htb]
\vspace{15mm}
\epsfig{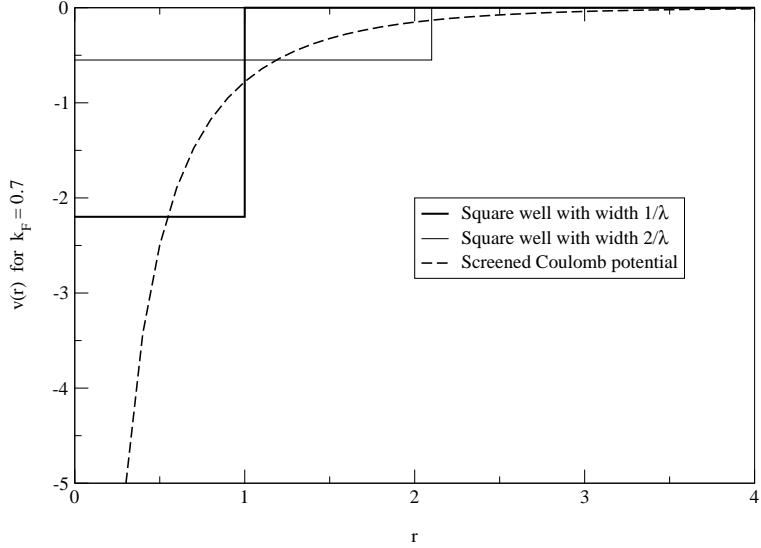}
\caption[]{$v(r)$ for the three potentials plotted as a function of $r$.}
\label{Vversusr}
\end{figure}

In order to get some more insight in the $k_{\rm F}$ dependences
shown in FIG. \ref{PaZdcorr} we display the quantities
$Z_{\rm corr}^{0}$, $Z_i(\rm pot)$ and $Z_{\rm F}$ in FIG. \ref{PaZiZ0corr}.
It is seen that all curves have a positive slope, apart from the
one for $Z_{\rm F}$ for the weaker $r_0=1/\lambda$ potential. This
is certainly related to the seemingly deviant
$Z_d/Z_{\rm F}$ curve in FIG. \ref{PaZdcorr} and the fact that the
$Z_d/Z_{\rm F}$ curves have been obtained by an accurate numerical
evaluation of Eq. (\ref{Sorb7opform1}), while the $Z_{\rm corr}/Z_{\rm corr}^{0}$
results 'suffer' from the $|q>\rightarrow|\psi_k>$ replacement. The
effect of this replacement is larger for the weaker potential, which
can be seen from both $Z_{\rm corr}/Z_{\rm corr}^{0}$ curves in
FIG. \ref{PaZdcorr} and the $Z_{\rm corr}^{0}$ curves in FIG. \ref{PaZiZ0corr}.
This is understandable if one realizes, that the effect of the
potential on the wave functions increases with its strength, while
the effect of the boundaries remains unchanged.
Further we observe that $Z_{\rm corr}^{0}$, $Z_i(\rm pot)$ and $Z_{\rm F}$
lead to different curves. The difference between $Z_{\rm corr}^{0}$ and
$Z_i(\rm pot)$ as shown in FIG. \ref{PaZiZ0corr} is certainly due to
the $|q>\rightarrow|\psi_k>$ replacement. However, Eq. (\ref{Zcorr0})
implies that for the screened Coulomb potential one should find
$Z_i(\rm pot)=Z_{\rm corr}^{0}=Z_i=Z_{\rm F}=1$.
This means that a difference between $Z_i(\rm pot)$ and $Z_{\rm F}$
uncovers some limitation of the use of the model square
well potential. Although the differences in shape of the three
potentials are known, in FIG. \ref{Vversusr} we show their shapes
for a certain $k_{\rm F}$ value, for which we chose a value 
in the middle of 0.7.

\section{The force expression of Bosvieux and Friedel}
\label{friedelfout}

The starting expression of Bosvieux and Friedel for the driving force is
\begin{equation}
\label{ForceBF}
{\bf F}_{T=0}=-<\Psi|\nabla_{{\bf R}_1}(V_{ei}+\delta V)|\Psi>,
\end{equation}
in which the state $|\Psi>$ is a solution of the Schr\"{o}dinger equation
for the system in the presence of an applied field. This means that
one has to solve the time dependent Schr\"{o}dinger equation
\begin{equation} 
\label{SchroedBF}
i\frac{\partial\Psi(t)}{\partial t}={\cal H}(t)\Psi(t)\equiv
\big(H+\delta V(t)\big)\Psi(t).
\end{equation}
The subscript $T=0$ is added by the present author in order to distinguish
this force from the force given in Eq. (\ref{Windplus}).
The derivation of Eq. (\ref{Windplus}), in which a solution
of the Liouville equation is used, has been given in the literature
many times. Because the approach through the system wave function is
typical for the theory of Bosvieux and Friedel, and because some
questions can be raised about their solution, we give crucial steps
of the derivation. We solve Eq. (\ref{SchroedBF}) by using the
interaction representation for $\Psi(t)$, defined as
\begin{equation} 
\label{PsiDirac} 
\Psi_I(t)\equiv e^{iHt}\Psi(t).
\end{equation}
The equation for $\Psi_I(t)$ becomes
\begin{equation} 
\label{EqPsiDirac}
i\frac{\partial\Psi_I(t)}{\partial t}=e^{iHt}\delta V(t)e^{-iHt}\Psi_I(t).
\end{equation}
After integrating this equation and using that for $t\rightarrow -\infty$
the system is in the ground state of the unperturbed system Hamiltionian $H$,
one finds for $\Psi(t)$ linearly in $\delta V$
\begin{equation} 
\label{Psiself}
\Psi(t)=-ie^{-iHt}\int_{-\infty}^t dt' e^{iHt'}\delta V(t')e^{-iHt'}
\Psi_I(-\infty)+e^{-iHt}\Psi_I(-\infty).
\end{equation}
With $\delta V(t)=\delta V e^{at}$, applying the
substitution $t-t'\equiv s$, and considering an arbitrary time in the
present, so $t=0$, this becomes
\begin{equation}
\label{Psiself3}
\Psi(0)\equiv\Psi=-i\int_0^\infty dt e^{-(iH+a)t}\delta V e^{iHt}
\Psi_I(-\infty)+\Psi_I(-\infty).
\end{equation}
If one calculates matrix elements with this $|\Psi>$, the factor
$e^{-iE_0\infty}$ in the state $|\Psi_I(-\infty)>$ drops out so that
just the ground state $|\psi_0>$ of $H$ remains.

By this one finds for Eq. (\ref{ForceBF})
\begin{eqnarray}
\label{ForceBF2}
{\bf F}_{T=0}&&=-<\psi_0|\nabla_{{\bf R}_1}(\delta V)|\psi_0>+
i\int_0^\infty dt~e^{-at}<\psi_0|(\nabla_{{\bf R}_1}V_{ei}) e^{-iHt}\delta V e^{iHt}
|\psi_0> +~~{\rm c.c.}\nonumber\\&&
=Z_ie{\bf E}+i\int_0^\infty dt~e^{-at}<\psi_0|
\Big[(\nabla_{{\bf R}_1}V_{ei}), e^{-iHt}\delta Ve^{iHt}\Big]|\psi_0>.
\end{eqnarray}
Because $H$ commutes with the
coordinates ${\bf R}_\alpha$ in $\delta V$ only the electron coordinates
survive. Using the definition of the force operator, Eq. (\ref{FopdVdR}), and
the hermitian property of $H$ it is clear that the time dependence
can be applied to ${\bf F}_{\rm op}$ as well, and Eq. (\ref{ForceBF2}) can be written as
\begin{equation} 
\label{ForceBF3}
{\bf F}_{T=0}=Z_ie{\bf E}-ie\int_0^\infty dt~e^{-at}<\psi_0| 
\Big[{\bf F}_{\rm op}(t),{\bf E}\cdot\sum_j{\bf r}_j\Big]|\psi_0>.
\end{equation}
Interestingly, Eq. (\ref{ForceBF3}) is precisely the zero-temperature equivalent of
Eq. (\ref{Windplus}). This becomes even more clear if one writes down the form
which shows up after the reduction of Eq. (\ref{ForceBF3}) to
single particle states denoted by $|q>$.
\begin{equation}
\label{ForceBFsp}
{\bf F}_{T=0}=Z_ie{\bf E}-ie\int_0^\infty dt~e^{-at}\sum_q
<q|\Big[{\bf f}^1(t),{\bf E}\cdot{\bf r}\Big]|q>.
\end{equation}
The force operator ${\bf f}^1$ is defined in Eq. (\ref{FopdVdR}).
At $T=0$ the sum over the single particle states has a sharp
cut-off at $\epsilon_q=\epsilon_{\rm F}$. The finite temperature
equivalent of Eq. (\ref{ForceBFsp}) can be written as
\begin{equation}
\label{ForceBFspFT}
{\bf F}=
Z_ie{\bf E}-ie\int_0^\infty dt~e^{-at}tr \bigg\{n(h)\Big[{\bf f}^1(t),
{\bf E}\cdot{\bf r}\Big]\bigg\},
\end{equation}
in which the Fermi-Dirac distribution $n(\epsilon)$ has been
inserted, see Eq. (\ref{nhop}).
Clearly, Eq. (\ref{ForceBFspFT}) is completely equivalent to
Eq. (\ref{Windred}) of the present text. By this electromigration
theory can be considered as being unified. Apparently, the starting
formula of Bosvieux and Friedel was correct, but these authors
did not recognize its precise contents. In fact, they wrote down
surface-integral terms, by this not appreciating the hermitian
property of the system Hamiltonian. This property implies that these terms
are zero, but their full-screening results were
derived from these terms. In addition, they missed the
power of their starting formula, Eq. (\ref{ForceBF}), by taking the
$a\rightarrow 0$ limit in too early a stage of the derivation.

\section{Concluding remarks and perspectives}
\label{conclusions}

The amount of screening of the direct force on a proton in an
electric-current carrying metal has been shown to lie between
5 and 25\%. By this the full-screening prediction of Bosvieux and
Friedel has been invalidated, completely in agreement with an
earlier result obtained by Sorbello.\cite{friedel,sor85} On top of
that, the surface integral terms used by Bosvieux and Friedel
to derive their full-screening result
appear to be zero, due to the hermitian property of the Hamiltonian.
Interestingly, it has been shown explicitly that the starting expression
of Bosvieux and Friedel for the driving force is the
zero-temperature limit of all linear-response expressions
used in the literature since their introduction by Kumar
and Sorbello.\cite{kumar}

All calculations up to now use a jellium model for a metal, or
are not applicable to transition metals.\cite{ishida}
In view of the description presented it becomes feasible
to account for real metallic effects. These effects
have been accounted for for the wind force to a large detail,
but for the direct force this was too much involved up
to now.\cite{sor97} Such a development would be interesting, because
this may lead to an explanation of a measured
result which has not been understood yet, up to now. For most hydrides a
direct valency for the hydrogen has been measured which is of the order of unity.
However, in Nb(H) a direct valency was found of about 0.44.
Such a deviating value may arise from multiple scattering effects of
the electrons around a proton surrounded by metallic atoms, which
can be accounted for in a finite-cluster description. It is
worth while to investigate this possibility, because in the development
of the description of the wind force surprising positive values for the wind valence
in V(H) and Nb(H) were found, which were in agreement with experiment \cite{brand}.
The surprise comes from the fact, that in a system composed of a finite cluster
embedded in a jellium the electron dispersion relation is still
free-electron like, from which one would expect a negative wind valence.
The calculated result must be due to the rather strong multiple
scattering effects, which were accounted for explicitly.
Similarly, it is worth while to develop a finite cluster
description for the direct valency, which is a straight
generalisation of the impurity in a jellium description
implemented so far. This is a feasible development if one
uses the simplified treatment presented above, of which
it has been shown that the expressions can be evaluated
in terms of the scattering phase shifts of the constituent
atoms. A slightly more explicit hint can be found elsewhere.\cite{lodkrakow}

{\bf Acknowledgement} I want to thank Jacques Friedel for an interesting
conversation and an extended correspondence over the recent months.

\appendix

\section{Crucial steps in the derivation of Eq. (\ref{ZcorrPhase})}
\label{crucial}

For the evaluation of $Z_{\rm corr}$ defined in Eq. (\ref{correction})
one needs the momentum matrix element ${\bf p}_{q'q}$ and the
matrix element of the commutator with the statistical distributions.
For ${\bf p}_{q'q}$ one writes
\begin{equation}
\label{pvecqaccq}
{\bf p}_{q'q}=\frac{(4\pi)^2}{8\pi^3}\sum_{LL'}
i^{\ell-\ell'}Y_{L'}(\hat{k}')Y_{L}^\ast(\hat{k})\int d^3r
R_{L'}^\ast({\bf r}){\bf p}R_L({\bf r}),
\end{equation}
in which Eq. (\ref{psikrL}) with $\Omega\rightarrow 8\pi^3$
has been used for the wave functions
and $R_L({\bf r})\equiv R_\ell(r,k)Y_L(\hat{r})$.
If one represents the scattering potential by a square well with depth
$v_0$ the inner radial solution is a Bessel function as well, so
$R_{\ell}(r,k)=A_{\ell}j_{\ell}(k_vr)$, with $k_v=\sqrt{k^2+v_0}$, and
one finds
\begin{equation}
\label{YLpRl}
\int Y_{L'}^\ast(\hat{r}){\bf p}~R_L({\bf r})d\hat{r}=
i^{\ell'-\ell}k_v{\bf D}_{L'L}R_{\ell'\ell}(r,k),
\end{equation}
in which the equality
\begin{equation}
\label{YLpjl2}
\int Y_{L'}^\ast(\hat{r}){\bf p}~j_L({\bf r})d\hat{r}=
i^{\ell'-\ell}k~{\bf D}_{L'L}j_{\ell'}(kr)
\end{equation}
has been used and
\begin{equation} 
\label{DLL}
{\bf D}_{L'L}\equiv\int d\hat{k}Y_{L'}^\ast(\hat{k})\hat{k}Y_{L}(\hat{k}).
\end{equation}
The double $\ell$-label in Eq. (\ref{YLpRl}) refers to the fact,
that the factor $A_{\ell}$ is not changed by the momentum operation and
the angular integration, so that
\begin{equation}
\label{Rellellaccell}
k_vR_{\ell'\ell}(r,k)=k_v\frac{A_{\ell}}{A_{\ell'}}R_{\ell'}(r,k)\rightarrow
k\left(j_{\ell'}(kr)-ikt_{\ell}h_{\ell'}^+(kr)\right).
\end{equation}
If one substitutes Eq. (\ref{YLpRl}) in Eq. (\ref{pvecqaccq}) one obtains
\begin{equation}
\label{pvecqaccq2}
{\bf p}_{q'q}=\frac{2}{\pi}\sum_{LL'}
Y_{L'}(\hat{k}')Y_{L}^\ast(\hat{k}){\bf D}_{L'L}I_{\ell'\ell}^{RR}(k',k)
\end{equation}
with
\begin{equation}
\label{IRRell}
I_{\ell'\ell}^{RR}(k',k)=\int_0^\infty r^2drk_rR_{\ell'}^\ast(r,k')R_{\ell'\ell}(r,k),
\end{equation}
and in which $k_r=k_v$ inside the range of the potential and
$k_r=k=k_{\rm F}$ outside of it. Using the equality
\begin{equation}
\label{intjkjkacc}
\int_0^\infty r^2drj_\ell(k'r)j_\ell(kr)=\frac{\pi\delta(k-k')}{2k^2},
\end{equation}
it appears to be possible to reduce the integral $I_{\ell'\ell}^{RR}(k',k)$ to
\begin{equation}
\label{IRRellkkacc2}
I_{\ell'\ell}^{RR}(k',k)=
\frac{k\pi}{2}\frac{\delta(k-k')}{k^2}
\bar{I}_{\ell'\ell}^{RR},
\end{equation}
in which
\begin{equation}
\label{IRRbar}
\bar{I}_{\ell'\ell}^{RR}\equiv 1-ikt_{\ell}+ikt_{\ell'}^\ast-
2ikt_{\ell}ikt_{\ell'}^\ast.
\end{equation}

Now we turn to the other matrix element in Eq. (\ref{correction}).
Using Eq. (\ref{nhmnh0}) one finds
\begin{equation}
\label{new1}
<q|[{\bf r},n(h)-n(h_0)]|q'>=-\int_0^{\beta}ds<q|\bigg[{\bf r},n(h)e^{sh}v
e^{-sh_0}(1-n(h_0))\bigg]|q'>.
\end{equation}
We remind the reader that
in this equation $h$ and $v$ just refer to the system with one
impurity, so to $h^1$ and $v^1$. Now we use the equality
(\ref{lambdaform}) and the following related equality
\begin{equation}
\label{ebhrnu}
[e^{\beta h},{\bf r}]=-\frac{i}{m}\int_0^\beta ds~{\bf p}(s)e^{\beta h}
\end{equation}
twice, one time for $h$ and one time for $h_0$. By that the commutator in the
right hand side of Eq. (\ref{new1}) can be written as
\begin{eqnarray}
\label{new11}
&&\bigg[{\bf r},n(h)e^{sh}ve^{-sh_0}(1-n(h_0))\bigg]=-\frac{i}{m}n(h)\Big(
\int_0^{\beta}ds'{\bf p}(s')(1-n(h))e^{sh}ve^{-sh_0}\nonumber\\&&
-\int_0^sds'{\bf p}(s')e^{sh}ve^{-sh_0}+e^{sh}ve^{-sh_0}{\bf p}~\big(s-
\beta n(h_0)\big)\Big)\Big(1-n(h_0)\Big).
\end{eqnarray}
It will be clear that ${\bf p}(s)$ in the first and second term refers to $h$.
 
Now we develop the $qq'$ matrix element of this operator as it occurs
in Eq. (\ref{new1}). In that we will make use of the property proven above
through Eq. (\ref{pvecqaccq2}) with (\ref{IRRellkkacc2}), namely that the energies
$\epsilon_{q}$ and $\epsilon_{q'}$ in the matrix element
${\bf p}_{qq'}$ are equal, and of the equality of the energies
$\epsilon_{k}$ and $\epsilon_{q}$ in the overlap $<k|q>$,
see Eq. (\ref{kacpsik}). In the
first and second term we have to insert two complete sets, one $q$ set
$|q'''><q'''|$ and one $k$ set $|k''><k''|$. In the third and
fourth term one needs the complete $k$ set only. This way one writes
for the $qq'$ matrix element in the left hand side of Eq. (\ref{new1})
\begin{equation}
\label{new2}
<q|[{\bf r},n(h)-n(h_0)]|q'>=\frac{i\beta^2}{m}n_q
\bigg[\sum_{q'''}\sum_{k''}{\bf p}_{qq'''}
(\frac{1}{2}-n_{q'''})t_{k''k'''}^\ast+
\sum_{k''}t_{k''k}^\ast {\bf k}^{''}
(\frac{1}{2}-n_{k''})\bigg](1-n_{k''})<k''|q'>,
\end{equation}
in which the potential matrix element $v_{qk'}=v_{k'q}^\ast$, between
an exact scattered state
$|q>=|\psi_k>$, see Eq. (\ref{psikrL}), and an unperturbed
state, a plane wave $|k'>$, has been replaced by the corresponding
{\it t} matrix element $t_{k'k}^\ast$.
 
By now we have developed the means for bringing $Z_{\rm corr}$
in a manageable form. One has to take the inner product of
the matrix element given by Eq. (\ref{new2}) with $\frac{i}{3}{\bf p}_{q'q}$
and to carry out the summations. Because the summations are equivalent to
integrals and the absolute values of $all$ $k$-vectors involved
are equal through delta functions, one just has to carry out the
angular integrations. We write for $Z_{\rm corr}$ in Eq. (\ref{correction})
\begin{equation}
\label{Znumerred}
Z_{\rm corr}=\frac{i}{3}\int d\hat{k}\int d\hat{k'}\int d\hat{k''}\left(\int d\hat{k'''}
\right)<q|[{\bf r},n(h^1)-n(h_0)]|q'>\cdot{\bf p}_{q'q},
\end{equation}
in which the right hand side of Eq. (\ref{new2}) is supposed to
have been substituted. The angular integration over $\hat{k'''}$
applies to the first term in the right hand side of Eq. (\ref{new2}) only.
The product of statistical factors which shows up can be simplified
and in the $T\rightarrow 0$ limit be written as follows
\begin{equation}
\label{statfactors}
-\beta^2n_k(1-n_k)(\frac{1}{2}-n_k)=\frac{1}{2}\beta\frac{\partial}
{\partial\epsilon_k}\big(n_k(1-n_k)\big)\rightarrow\frac{1}{2}\frac{\partial}
{\partial\epsilon_k}\delta(\epsilon_k-\epsilon_{\rm F}).
\end{equation}
If one substitutes this equality, uses Eq. (\ref{pvecqaccq2}) with
Eqs. (\ref{IRRellkkacc2}) and (\ref{IRRbar}), and Eq. (\ref{new2}),
accounts for the factor of 2
due to the spin degeneracy and carries out all angular
integrations, one finds for $Z_{\rm corr}$
of Eq. (\ref{Znumerred})
\begin{eqnarray}
\label{Znumerred2}
Z_{\rm corr}=&&\frac{2}{3\pi m}\int_0^\infty k^2dk
\frac{\partial}{\partial\epsilon_k}
\delta(\epsilon_k-\epsilon_{\rm F})\sum_{LL'}{\bf D}_{LL'}\cdot{\bf D}_{L'L}
(1-ikt_{\ell'})\times\nonumber\\&&
\left(t_{\ell'}^\ast\bar{I}_{\ell\ell'}^{RR}
+t_{\ell}^\ast\right)k^2\bar{I}_{\ell'\ell}^{RR}=
 -\frac{4}{3\pi m}
\int_0^\infty k^3dk\Big(\frac{\partial}{\partial\epsilon_k}
\delta(\epsilon_k-\epsilon_{\rm F})\Big)F(\epsilon_k),
\end{eqnarray}
in which has been used that
\begin{equation}
\label{Ibarell}
\bar{I}_{\ell\ell'}^{RR}=\cos(\delta_{\ell}-\delta_{\ell'})~
e^{-i(\delta_{\ell}-\delta_{\ell'})},
\end{equation}
and that
\begin{equation}
\label{DnuDmu}
\sum_{mm'}D_{LL'}^\nu{D}_{L'L}^\mu =\frac{1}{3}\delta_{\mu\nu}
\bigg((\ell+1)~\delta_{\ell',\ell+1}+\ell~\delta_{\ell',\ell-1}\bigg).
\end{equation}

In the right hand side of Eq. (\ref{Znumerred2}) one recognizes
the second member of Eq. (\ref{ZcorrPhase}), by which the derivation has
been completed. The derivations leading to Eqs. (\ref{ZcorrPhase01}),
(\ref{ZcorrPhase02}) and (\ref{SorbScatt}) are similar, and they are simpler as well.

\end{document}